\documentclass[letterpaper,12pt]{article}   
\usepackage{osajnl2, amsfonts} 

\begin{document}

\title{Analytical modeling and 3D finite element simulation 
of line edge roughness in scatterometry} 


\author{Akiko Kato,$^{1,*}$ Sven Burger,$^{2,3}$ and Frank Scholze$^{4}$}
\address{$^1$Hochschule Ruhr West - University of Applied Sciences,\\D\"umptener
  Stra{\ss}e 45, 45476 M\"ulheim an der Ruhr, Germany}
\address{$^2$Zuse Institute Berlin (ZIB), Takustra{\ss}e 7, 14195 Berlin, Germany}
\address{$^3$JCMwave GmbH, Bolivarallee 22, 14050 Berlin, Germany}
\address{$^4$Physikalisch-Technische Bundesanstalt (PTB), Abbestra{\ss}e 2-12, 10587 Berlin, Germany}
\address{$^*$Corresponding author: akiko.kato@hs-ruhrwest.de}

\begin{abstract}
The influence of edge roughness in angle resolved scatterometry at periodically structured surfaces is investigated. A good description of the radiation interaction with structured surfaces is crucial for the understanding of optical imaging processes like, e.g. in photolithography. We compared an analytical 2D model and a numerical 3D simulation with respect to the characterization of 2D diffraction of a line grating involving structure roughness. The results show a remarkably high agreement. The diffraction intensities of a rough structure can therefore be estimated using the numerical simulation result of an undisturbed structure and an analytically derived correction function. 
This work allows to improve scatterometric results for the case of practically relevant 2D structures. 
\end{abstract}


\maketitle 

This paper will be published in Applied Optics and is made available as an electronic preprint 
with the permission of OSA. 
Systematic or multiple reproduction or distribution to multiple locations via electronic 
or other means is prohibited.

\section{Introduction}

Scatterometry is a common technique for the characterization of nano-structured surfaces. It is an indirect measuring method inferring the properties of the scattering object from the light diffracted. The interaction between the scattering object and the incoming electromagnetic radiation is simulated numerically. Applying an optimization algorithm, the measured diffraction intensities are fitted to the calculated ones to derive the sought properties of the sample. It should be noted that the same numerical models are also used to predict, e.g. the imaging properties of mask structures in a photolithographic process. A better understanding of the accuracy of the optical models is therefore crucial also for many other applications than scatterometry. Scatterometry, however, is the natural environment for the investigation of the accuracy of optical models because it is explicitly based on the correlation of geometrical structure and optical properties.
Currently, scatterometry is used as a relative metrology method for process control and process development. Many attempts have been made to establish scatterometry as a traceable and absolute metrological method for dimensional measurements of nanostructured surfaces \cite{germer07a, germer07, naulleau05, gross09a}, some of them including the evaluation of measurements on test structures for line roughness \cite{boher05, wang07}. 

At the Physikalisch-Technische Bundesanstalt (PTB), detailed scatterometric investigations of an EUV photomask with periodic absorber line grating test patterns have been performed \cite{scholze08, bodermann09}. It was feasible to derive the information on the line profile by means of rigorous numerical modelling \cite{gross09a, pomplun08, burger11, scholze11}. A detailed analysis of the uncertainty contributions in the structure reconstruction process \cite{gross09a} revealed that structure disturbances such as line edge or line width roughness have significant impact on the angular distribution of the diffraction intensities \cite{scholze11, kato11}. Uncertainties in the actual intensity measurements are less important.
In \cite{kato10}, a Debye-Waller-type attenuation factor was found using a stochastic model of line roughness. It decreases the expectation value of the diffraction intensities perpendicular to the absorber lines. The calculations are based upon an analytical model describing the far-field diffraction of a disturbed 1D binary grating. The Debye-Waller factor found there was applied to the scattering of a patterned EUV multilayer photomask, and the roughness values found showed a high correlation compared to the rms roughness measured by scanning electron microscopy \cite{kato11}. Within a Monte Carlo simulation of the randomly disturbed EUV line grating over several periods, the exponential behaviour of the efficiency attenuation for the in-plane scattering has recently been confirmed \cite{gross11}. 
However, this recent treatment was for a 1D structure, which inherently only describes in-plane scatter between the diffraction orders. In realistic investigations of roughness in scatterometry \cite{boher05, wang07, schuster08} also out-of-plane scatter has to be included. It will be shown that in the Fraunhofer far-field approximation, the impact of the line roughness is given by a multiplicative term also for 2D structures. While scatterometry basically is the observation of the far-field, the condition for the Fraunhofer and Kirchhoff approximations of the structures being substantially larger than the wavelength is not generally met, particularly not for optical scatterometry at state-of-the-art semiconductor structures. This condition is equivalent to the existence of several orders of diffraction. For the example presented here, EUV scatterometry at structures of about 100 nm critical dimensions, it is met. Even for structures smaller than the wavelength, the spatial wavelengths of the perturbations (e.g. stitching errors) may be larger than the wavelength and roughness induced scatter as described by the formalism presented here is observable. For small structures and small spatial wavelength perturbations, however, no other diffraction orders than the zeroth order exist and other descriptions like the effective layer model \cite{schuster08, Bergner10} must be used to account for roughness.

For periodic edge roughness, the out-of-plane diffraction is described by Bessel functions. This approach can be generalized to random roughness by using a Fourier expansion of the rough contours. For the in-plane scatter, the correction factor converges for small roughness amplitudes to an exponential factor, as derived before in the 1D case. The analytical approach presented here is also compared with rigorous numerical calculations.
The EUV photomask used as an example consists of a quarter inch thick substrate with a reflective Mo/Si multilayer coating. The multilayer coating is terminated with a protective Si/SiO$_{2}$ capping layer. On top of the multilayer, an absorber stack is deposited, which is then structured by e-beam lithography. The paper is structured as follows: in section \ref{section_analytisch} an analytical study of the 2D line roughness is presented. In order to evaluate the scatterometry measurements we use a FEM based Maxwell solver as simulation tool. In section \ref{sectionsimulation} this numerical approach is applied to analyze the scattering caused by line roughness. In section \ref{bilderdeutungundinterpretation} the results of the two previous sections are compared and discussed. 

\section{Analytical modeling}\label{section_analytisch}

Within the classical Kirchhoff's diffraction theory, the electromagnetic field is treated as a scalar quantity. 
The grating is considered as an aperture with periodic slits which is irradiated by monochromatic radiation. 
Since the example we are presenting here is a grating consisting of periodic absorber lines on a mirror surface, 
we do not use transmission but reflection. Thus the grating is represented by a periodically varied reflectivity $r$ 
on the sample's surface in the $(x,y)$ plane.

\begin{equation} \label{r_undisturbed}
r(x,y)=\sum_{j=-n}^{n}\delta\left(x-jd
\right)\ast\mathrm{rect}\left(\frac{x}{c}\right)
\qquad \textrm{for all} \qquad x\in\left[-\frac{Nd}{2},\frac{Nd}{2}\right], y\in\mathbb{R}\,.
\end{equation}

\begin{figure}[htbp]
\begin{center}
\includegraphics[width=\textwidth]{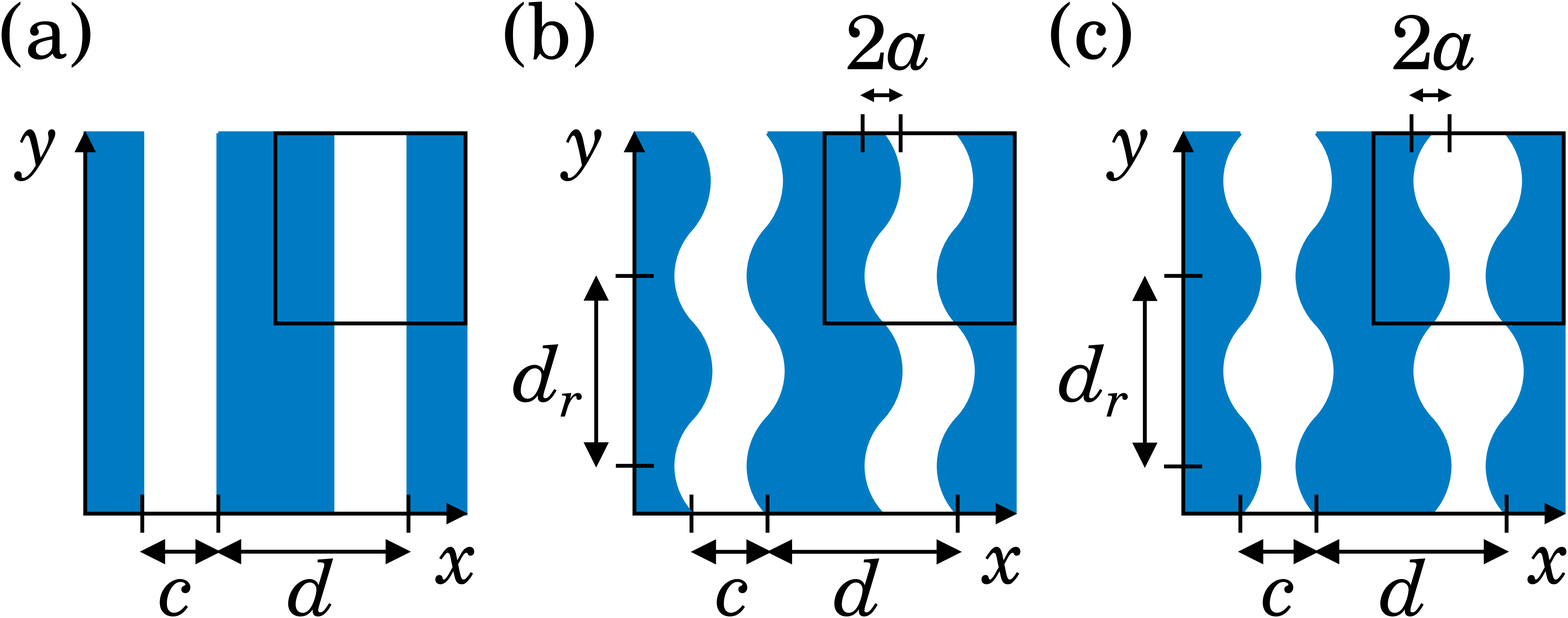}
\end{center}
\caption{\label{2Dmodel} 2D binary grating without roughness (a).    
Deterministic `roughness' models, line edge roughness (b) and line width roughness (c). The framed boxes indicate the unit cells of the numerical simulation, cf. section \ref{sectionsimulation}.}
\end{figure}

Here, ``$\ast$'' stands for the convolution over $x$. The lines are parallel to the $y$-axis, cf. Fig. \ref{2Dmodel} (a).
$d$ is the period of the structure in $x$-direction, $c$ is the width of the reflecting area. For simplicity, we set the reflectivity value of the open reflecting multilayer to 1 and that of the absorber lines to zero, and it is assumed that the length of the grating is infinite in $y$-direction. $n\in\mathbb{N}$, and $N=2n+1$ gives the total number of periods. Henceforth, we will refer to this kind of grating, which is invariant in $y$, as an undisturbed grating and the subscript $r$ denotes the quantities referring to the undisturbed grating. 
In the far field, the amplitude $E$ of the electric field is proportional to the Fourier transform of the reflection function $r$ given by Eq. (\ref{r_undisturbed}),

\begin{equation}\label{i_0}
E\left(k_{x},k_{y}\right)\propto\mathcal{F}\left\{r\right\}\left(k_{x},k_{y}\right)\,,
\end{equation}

where $\mathcal{F}\left\{r\right\}$ stands for the Fourier transform of $r$ with the wave vectors $\left(k_{x},k_{y}\right)$ as arguments.
In the scalar approximation, the intensity $I_{r}$ is given by the square of the electrical field:

\begin{equation}\label{i_0intensitaet}
I_{r}\left(k_{x},k_{y}\right)=\left|E\left(k_{x},k_{y}\right)\right|^{2}\propto
\left(c\,\frac{\sin\frac{k_{x}c}{2}}{\frac{k_{x}c}{2}}\,
\frac{\sin\frac{Nk_{x}d}{2}}{\sin\frac{k_{x}d}{2}}\,2\pi\delta\left(k_{y}\right)\right)^{2}\,.
\end{equation}

Eq. (\ref{i_0intensitaet}) describes the well-known Fraunhofer diffraction of a grating, whose
discrete diffraction orders are found at $k_{x}=2\pi m/d$ for all $m\in\mathbb{Z}$ and $k_{y}=0$. The corresponding intensity values will be called diffraction intensities of order $m$ in $x$, respectively, throughout the paper. $\delta$ is the Dirac delta function. We are aware of the simplicity of this binary grating model, neglecting the mask's 3D geometry, all material properties, and the reflection characteristics of the multilayer coating. Therefore, the form factor given by the sinc term in Eq. (\ref{i_0intensitaet}) will not be found exactly in measurements. The advantage is, however, that based on this model, the structure roughness can be treated analytically. In \cite{kato11, kato10} the impact of the stochastic 1D roughness was successfully described. The roughness could explain the observed differences in the side wall angle as reconstructed by angle-resolved scatterometry and measured independently by atomic force microscopy, whereas other possible uncertainty contributions could not explain this difference adequately \cite{gross09a}.

The roughness models we consider are depicted in Fig. \ref{2Dmodel} (b) and
(c). The line edges have  sinusoidal shapes that are in phase for all edges in
the case of line edge roughness (LER), keeping a fixed value for the line
width. In case of the line width roughness (LWR), the line edges are phase
shifted by $\pi$ which means that the line centre position is fixed along the
lines while the line width shows a periodic modulation. The framed boxes in
Fig. \ref{2Dmodel} indicate the unit cells calculated with the finite element
method (FEM) in section \ref{sectionsimulation} using periodic boundary
conditions. Note that any random roughness can be expressed by a Fourier
expansion as the sum of sinusoidal waves. It is therefore sufficient to find a
solution for the sinusoidal case and the general solution is just a linear
superposition. Therefore, this approach is widely used. E.g. \cite{schuster09}
presents the numerical modelling of 2D line edge roughness for optical
scatterometry using the field-stitching method in rigorous coupled-wave
analysis. In this paper, we will first derive an analytical description of the
line edge modulation and will then compare the results to 3D FEM simulations.

\subsection{Line edge roughness}

The reflectivity of a binary line-and-space grating with LER as outlined in Fig. \ref{2Dmodel} (b) is given by

\begin{equation}
\label{f_ler}
f(x,y)=\sum_{j=-n}^{n}\delta\left(x-\left(jd-a\cos\left(2\pi y/d_{r}\right)\right)\right)\ast\mathrm{rect}\left(\frac{x}{c}\right)
\quad \textrm{for all} \quad x\in\left[-\frac{Nd}{2},\frac{Nd}{2}\right], y\in\mathbb{R}\,,
\end{equation}

where ``$\ast$'' stands for the convolution over $x$ and $a\cos\left(2\pi y/d_{r}\right)$ desribes the sinosoidal edge with spatial period $d_{r}$ and amplitude $a$.
In the following equations $f$ stands for the LER-disturbed grating instead of $r$ for the undisturbed solution in (\ref{i_0intensitaet}).
Then the far field intensity can ben expressed by means of the Fourier transform of $f$,

\begin{equation}
I_{f}\left(k_{x},k_{y}\right)\propto\left|\mathcal{F}\left\{f\right\}\left(k_{x},k_{y}\right)\right|^{2}
=\left|c\,\frac{\sin\frac{k_{x}c}{2}}{\frac{k_{x}c}{2}}\,\frac{\sin\frac{k_{x}dN}{2}}{\sin\frac{k_{x}d}{2}}\sum_{m=-\infty}^{\infty}i^{m}J_{m}\left(ak_{x}\right)2\pi\delta\left(k_{y}-m2\pi/d_{r}\right)\right|^{2}\,.
\end{equation}

Here, the Jacobi-Anger identity \cite{colton98} has been used to express the Fourier transform in form of
Bessel functions $J_{m}$ of the first kind of order $m$.
For a given $k_{y}$, a diffraction order exists, if $\exists m\in\mathbb{Z} \quad k_{y}= m2\pi/d_{r}$. In this case,

\begin{equation}\label{ler}
I_{f}\left(k_{x},k_{y}\right)\propto
\left|c\,\frac{\sin\frac{k_{x}c}{2}}{\frac{k_{x}c}{2}}\,\frac{\sin\frac{k_{x}dN}{2}}{\sin\frac{k_{x}d}{2}}\,i^{m}J_{m}\left(ak_{x}\right)2\pi\right|^2\,.
\end{equation}

A comparison with Eq. (\ref{i_0}) yields

\begin{equation}\label{ler_quotient}
I_{f}\left(k_x,m2\pi/d_{r}\right)=J_{m}^2\left(ak_x\right)\times I_{r}\left(k_x,0\right)\,,
\end{equation}

a formula describing the intensity of the LER grating $f$ given by Eq. (\ref{f_ler}) with respect to the undisturbed reference grating $r$ given by Eq. (\ref{r_undisturbed}).
The equation states that the impact of line roughness does not depend on the roughness period $d_{r}$, but only on its amplitude $a$. Therefore, the result can be directly generalized to random roughness by Fourier expansion of the roughness contours. On the other hand it is sufficient to verify only the sinusoidal case, which is much easier to handle, with FEM. In the case of 1D random roughness \cite{kato10}, the attenuation factor of Debye-Waller type caused by roughness also was a function only of the standard deviation of the random line position or the random line width. In the ratio of the diffraction intensities $I_{f}(k_{x},k_{y})$ of the disturbed grating for finite values of $k_{y}$, to the diffraction intensities $I_{r}(k_{x},0)$ of the undisturbed grating, the term $I_{r}(k_{x},0)$ cancels completely, Eq. (\ref{ler_quotient}). The ratio only depends on the line roughness. Therefore, this ratio can be easily compared to measurements or numerical simulations, even if the values for $I_{r}$ do not well represent the measured or simulated values. We present a respective numerical comparison in section \ref{bilderdeutungundinterpretation}. 
Furthermore, for a given diffraction order $n$ in $x$, $k_{x}= n2\pi/{d}$ all diffraction orders $m$ in $y$ can be summed up:

\begin{eqnarray}
\sum_{\mathrm{observable~} m}I_{f}\left(k_x,m2\pi/d_{r}\right)&=&I_{r}\left(k_x,0\right)\sum_{\mathrm{observable~} m}J_{m}^2\left(ak_x\right)\nonumber\\
\label{energieerhaltung}
&\le&I_{r}\left(k_x,0\right)\sum_{m\in\mathbb{Z}}J_{m}^2\left(ak_x\right)=I_{r}\left(k_x,0\right)\,.
\end{eqnarray}

Here, ``observable $m$'' is a subset of $\mathbb{Z}$ referring to the diffraction orders which are observable, i.e. reflected back with scattering angles in $(-90^{\circ},90^{\circ})$ along the $y$-direction with respect to the surface normal.
The inequality (\ref{energieerhaltung}) states that the undisturbed intensity of each order in $x$ is spread through the line roughness over all orders in $y$-direction for this $k_x$ value. 

\subsection{Line width roughness}

The reflectivity of the absorber structure disturbed by LWR, as outlined in Fig.~\ref{2Dmodel}(c) is given by

\begin{equation}\label{g_lwr}
g(x,y)=\sum_{j=-n}^{n}\delta\left(x-jd\right)\ast\mathrm{rect}\left(\frac{x}{\mathrm{c}-2a\cos\left(2\pi y/d_{r}\right)}\right)
\quad \textrm{for all} \quad x\in\left[-\frac{Nd}{2},\frac{Nd}{2}\right], y\in\mathbb{R}\,.
\end{equation}
Using the trigonometric angle addition formula and the Jacobi-Anger identity, it follows that

\begin{eqnarray}
&&\mathcal{F}\left\{g\right\}\left(k_{x},k_{y}\right)\nonumber\\
&=&\sum_{j=-n}^{n}\int\limits_{-\infty}^{\infty}\int\limits_{-\infty}^{\infty}\delta\left(x-jd\right)\ast\mathrm{rect}\left(\frac{x}{\mathrm{c}-2a\cos\left(2\pi
    y/d_{r}\right)}\right)\exp\left(-ik_{x}x-ik_{y}y\right)\mathrm{d}x\mathrm{d}y\nonumber\\
&\stackrel{k_{x}\neq0}{=}&\sum_{j=-n}^{n}\int\limits_{-\infty}^{\infty}\exp\left(-ik_{x}jd\right)\frac{2}{k_{x}}\sin\frac{k_{x}\left(\mathrm{c}-2a\cos\left(2\pi
    y/d_{r}\right)\right)}{2}\exp\left(-ik_{y}y\right)\mathrm{d}y\nonumber\\
&=&\sum_{j=-n}^{n}\mathcal{F}\left\{\exp\left(-ik_{x}jd\right)\frac{2}{k_{x}}\sin\frac{k_{x}\left(c-2a\cos\left(2\pi y/d_{r}\right)\right)}{2}\right\}\left(k_{y}\right)\nonumber\\
&=&
c\,\frac{\sin\frac{k_{x}dN}{2}}{\sin\frac{k_{x}d}{2}}\left[\frac{\sin\frac{k_{x}c}{2}}{\frac{k_{x}c}{2}}\,J_{0}\left(k_{x}a\right)2\pi\delta\left(k_{y}\right)\right.\nonumber\\
&&\left.+\,\frac{\sin\frac{k_{x}c}{2}}{\frac{k_{x}c}{2}}\,2\sum_{n=1}^{\infty}(-1)^{n}J_{2n}\left(k_{x}a\right)\pi\left[\delta\left(k_{y}+2n2\pi/d_{r}\right)+\delta\left(k_{y}-2n2\pi/d_{r}\right)\right]\right.\nonumber\\
&&\left.+\,\frac{\cos\frac{k_{x}c}{2}}{\frac{k_{x}c}{2}}\,2\sum_{n=1}^{\infty}(-1)^{n}J_{2n-1}\left(k_{x}a\right)\pi\left[\delta\left(k_{y}+(2n-1)2\pi/d_{r}\right)+\delta\left(k_{y}-(2n-1)2\pi/d_{r}\right)\right]\right]\,.\nonumber\\
&&
\end{eqnarray}

Unlike as for LER in the previous section, here it is suitable to discuss the odd and the even diffraction orders in $y$ separately to further simplify the expressions.
Diffraction orders are found at each multiple of the roughness frequency $2\pi/d_{r}$ for $k_{y}$.
For the even orders $k_{y}$, \,$\exists m\in\mathbb{Z} \quad k_{y}=2m(2\pi/d_{r})$\,, we obtain:

\begin{equation}\label{lwrgeradeyordnungen}
\mathcal{F}\left\{g\right\}\left(k_{x},k_{y}\right)\propto 2\pi c\,
\frac{\sin\frac{k_{x}dN}{2}}{\sin\frac{k_{x}d}{2}}\,\frac{\sin\frac{k_{x}c}{2}}{\frac{k_{x}c}{2}}\,
(-1)^{|m|}J_{2|m|}\left(k_{x}a\right)\,.
\end{equation}

For the odd orders $k_{y}$, \,$\exists m\in\mathbb{Z} \quad k_{y}=(2m-1)2\pi/d_{r}$\, and \,$k_{x}\neq 0$\,, then

\begin{eqnarray}
\mathcal{F}\left\{g\right\}\left(k_{x},k_{y}\right)&\propto&2\pi c\,
\frac{\sin\frac{k_{x}dN}{2}}{\sin\frac{k_{x}d}{2}}\,\frac{\cos\frac{k_{x}c}{2}}{\frac{k_{x}c}{2}}\,
(-1)^{m}J_{2m-1}\left(k_{x}a\right) \qquad \textrm{for} \quad m>0\,,\\
\mathcal{F}\left\{g\right\}\left(k_{x},k_{y}\right)&\propto&2\pi c\,
\frac{\sin\frac{k_{x}dN}{2}}{\sin\frac{k_{x}d}{2}}\,\frac{\cos\frac{k_{x}c}{2}}{\frac{k_{x}c}{2}}\,
(-1)^{|m|+1}J_{2|m|+1}\left(k_{x}a\right) \qquad \textrm{for} \quad m\leq 0\,.
\end{eqnarray}

This implies that for the even orders

\begin{equation}\label{lwr_quotient1}
I_{g}\left(k_x,2m2\pi/d_{r}\right) = J_{2m}^2\left(ak_x\right)\times I_{r}\left(k_x,0\right)\,,
\end{equation}

where $I_{g}$ is the intensity distribution of the LWR grating given by Eq. (\ref{g_lwr}) and $I_{r}$ is the intensity of the undisturbed reference grating given by Eq. (\ref{r_undisturbed}).
For even orders in $y$, the ratio is therefore the same as was found in the case of line edge roughness, cf. Eq. (\ref{ler_quotient}). Note that for the zeroth order in $y$, the following approximation is found for small $ak_x$:

\begin{equation}
\frac{I_{f,g}\left(k_x,0\right)}{I_{r}\left(k_x,0\right)}=J_{0}^2\left(ak_x\right)\approx\exp\left(-\frac{\left(ak_x\right)^2}{2}\right)\,,
\end{equation}

for both LER and LWR. The exponential ratio was also found in the 1D stochastic case \cite{kato10}, where the deterministic amplitude $a$ is replaced by the standard deviation of the probability density for the line edge. It is worth mentioning that this exponential factor was also found independently for small angle X-ray scattering \cite{wang07}.
For the odd orders in $y$, we find the ratio

\begin{equation}\label{lwr_quotient2}
\frac{I_{g}\left(k_x,(2m-1)2\pi/d_{r}\right)}{I_{g}\left(k_x,(2p-1)2\pi/d_{r}\right)}=\frac{J_{2m-1}^2\left(ak_x\right)}{J_{2p-1}^2\left(ak_x\right)}\,,
\end{equation}

for $m,p\in\mathbb{N}$. For LWR, the form factor is no longer the sinc function. 
Therefore, the ratio to the undisturbed grating $r$ cannot be discussed as easily as for the even orders.
Because of the divergence in the odd orders in $y$, the case $k_{x}=0$ must be discussed separately. Setting $k_{x}=0$ in

\begin{equation}
\mathcal{F}\left\{f\right\}\left(k_{x},k_{y}\right)
=\sum_{j=-n}^{n}\mathcal{F}\left\{\exp\left(-ik_{x}jd\right)\left(c-2a\cos\left(2\pi
y/d_{r}\right)\right)\right\}\left(k_{y}\right)
\end{equation}

yields

\begin{equation}\label{lwr_kx_0}
\mathcal{F}\left\{f\right\}\left(0,k_{y}\right)
=2\pi N\left(c\,\delta\left(k_{y}\right)-a\left[\delta\left(k_{y}+2\pi/d_{r}\right)+\delta\left(k_{y}-2\pi/d_{r}\right)\right]\right)\,.
\end{equation}

Thus for $k_{x}=0$, diffraction orders in the $k_{y}$-direction exist only for $k_{y}=0$ or $k_{y}=\pm2\pi/d_{r}$ with relative amplitudes of $c/a$, respectively.

\section{Rigorous simulations of light scattering off EUV line masks with sinusoidal roughness}\label{sectionsimulation}

\begin{figure}[htbp]
\begin{center}
\includegraphics[width=\textwidth]{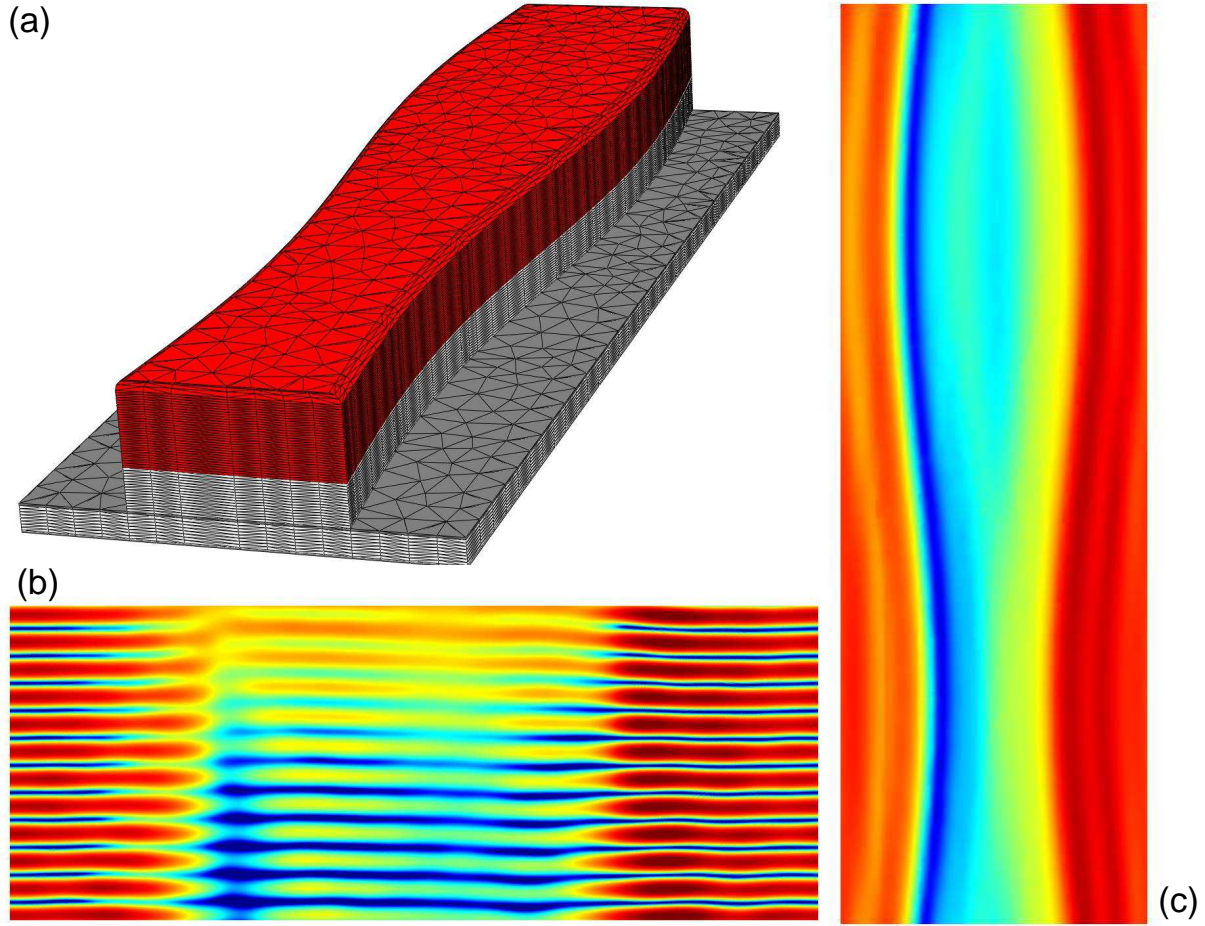}
\end{center}
\caption{\label{figure_fem_mesh}  (a) Visualization of parts of a FEM mesh discretizing an EUV line mask with sinusoidal LWR, amplitude $a=10$~nm and period $d_{r}=600$~nm (red: absorber, light gray: buffer). 
Mesh elements discretizing the surrounding vacuum and the multilayer mirror below the structure are not shown.
Mesh generated using the automatic mesh generator JCMgeo. 
(b) Pseudo-color visualization of the electromagnetic field intensity distribution $I(\vec{r}\,)$ on a logarithmic scale (colors blue to red
correspond to $\log(I) = -3\dots +1$) in a $x$-$z$ cross section through the upper part of the computational domain containing
the absorber structure.
(c) Same, in a $x$-$y$ cross section through the center of the absorber structure.}
\end{figure}

For rigorous numerical simulations of light scattering off EUV masks we use the time-harmonic FEM solver JCMsuite~\cite{pomplun07pssb}. The geometry of a unit cell as schematically depicted in Fig.~\ref{2Dmodel} is discretized with a 3D volume mesh. Fig.~\ref{figure_fem_mesh}\,(a) shows a visualization of a mesh for typical geometry parameters. Clearly the method also allows for more complex roughness models~\cite{lockau11pw}.

In the simulation, the structure is illuminated with S- and P-polarized plane waves (vacuum wavelength $\lambda_0=13.4$\,nm) at oblique incidence with an inclination angle of $6^{\circ}$. We performed simulations on 3D computational domains with a periodicity in $x$-direction, $d=200$\,nm, periodicities in $y$-direction, $d_r=200\,\dots 1000\,$nm, and a total thickness of absorber stack, Si/SiO$_{2}$ capping layer and Mo/Si multilayer coating (60 layers), $h_\mathrm{total}\approx 500$\,nm. The linewidth of the absorber structure was $\mathrm{c}= 100$\,nm, with sidewall angle $\alpha=90^{\circ}$ and top corner rounding radius $r_\mathrm{rounding}= 5$\,nm.  We show data for two different roughness periodicities in $y$-direction, $d_{r}=$300~nm and 600~nm, and for six different roughness amplitudes $a=n\times2.5~\textrm{nm}$, where $n=1,\dots,6$. 
To discretize the vectorial electric field solution to the Maxwell light scattering problem we use edge elements of $5^\mathrm{th}$ polynomial order.
An automatic, rigorous domain-decomposition method is used to separate the essentially 1D problem of light propagation in the multilayer mirror from the 3D problem in the part of the computational domain containing the 3D absorber structure~\cite{schaedle07dd}. 
Figs. \ref{figure_fem_mesh} (b), (c) show visualizations of the EUV light field intensity distribution in two 2D cross sections through the 3D computational domain. 

The amplitudes and phases of the reflected diffraction orders in the far field presentation are obtained from Fourier-transforms of the 3D near field solution. 
We note that the computational domain is relatively large, with a total volume of 10,000 to 50,000 cubic wavelengths. Therefore, the numerical accuracy of the 3D results was checked carefully in a previous convergence study and by comparison with a scatterometry experiment~\cite{burger11,burger11pm}. Numerical accuracies with relative errors lower than 0.01\% are reached for the central diffraction orders with relatively large total power; numerical accuracies with relative errors lower than 1\% are reached for higher diffraction orders with total intensities down to $10^{-6}$. 
The results presented here  are obtained using a standard multi(8)-processor computer: the typical memory usage (RAM) is in the range 10-100\,GB, computation times are in the range of minutes to tens of minutes.

\section{Comparison simulation vs. calculation}\label{bilderdeutungundinterpretation}

\begin{figure}[htbp]
\begin{center}
\includegraphics[width=\textwidth]{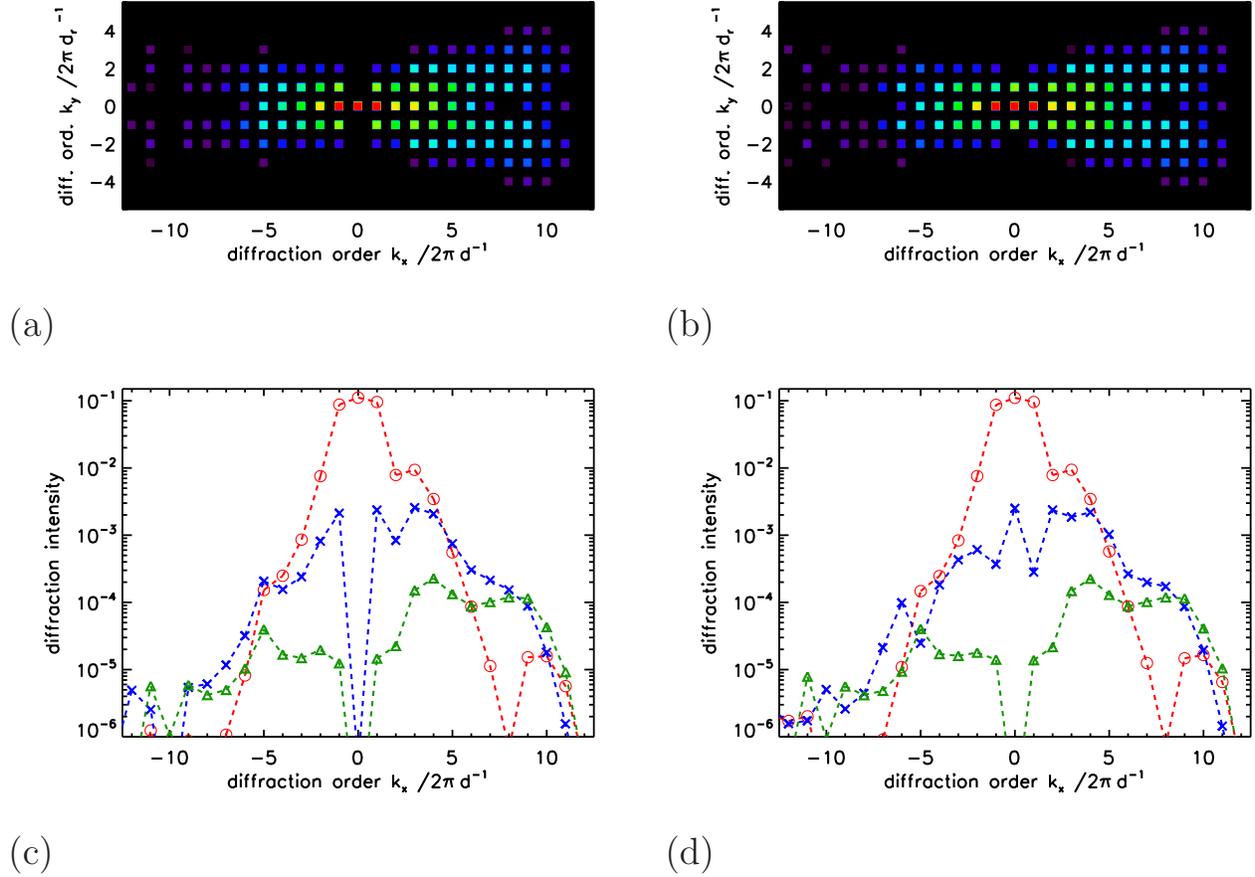}
\end{center}
\caption{\label{3dfem}  (a) and (b) Pseudo-color visualization of the
result of FEM simulations for the structure from Fig. \ref{figure_fem_mesh} for LER (left) and LWR (right) on a logarithmic scale (colors blue to red correspond to $\log(I) = -5\dots -1$). Numerical values are shown in (c) and (d) for the $0^{\mathrm{th}}$ (red circles), $1^{\mathrm{st}}$ (blue crosses), and $2^{\mathrm{nd}}$ (green triangles) order in $y$ of Figures (a) and (b), respectively.}
\end{figure}

\begin{figure}[htbp]
\begin{center}
\includegraphics[width=\textwidth]{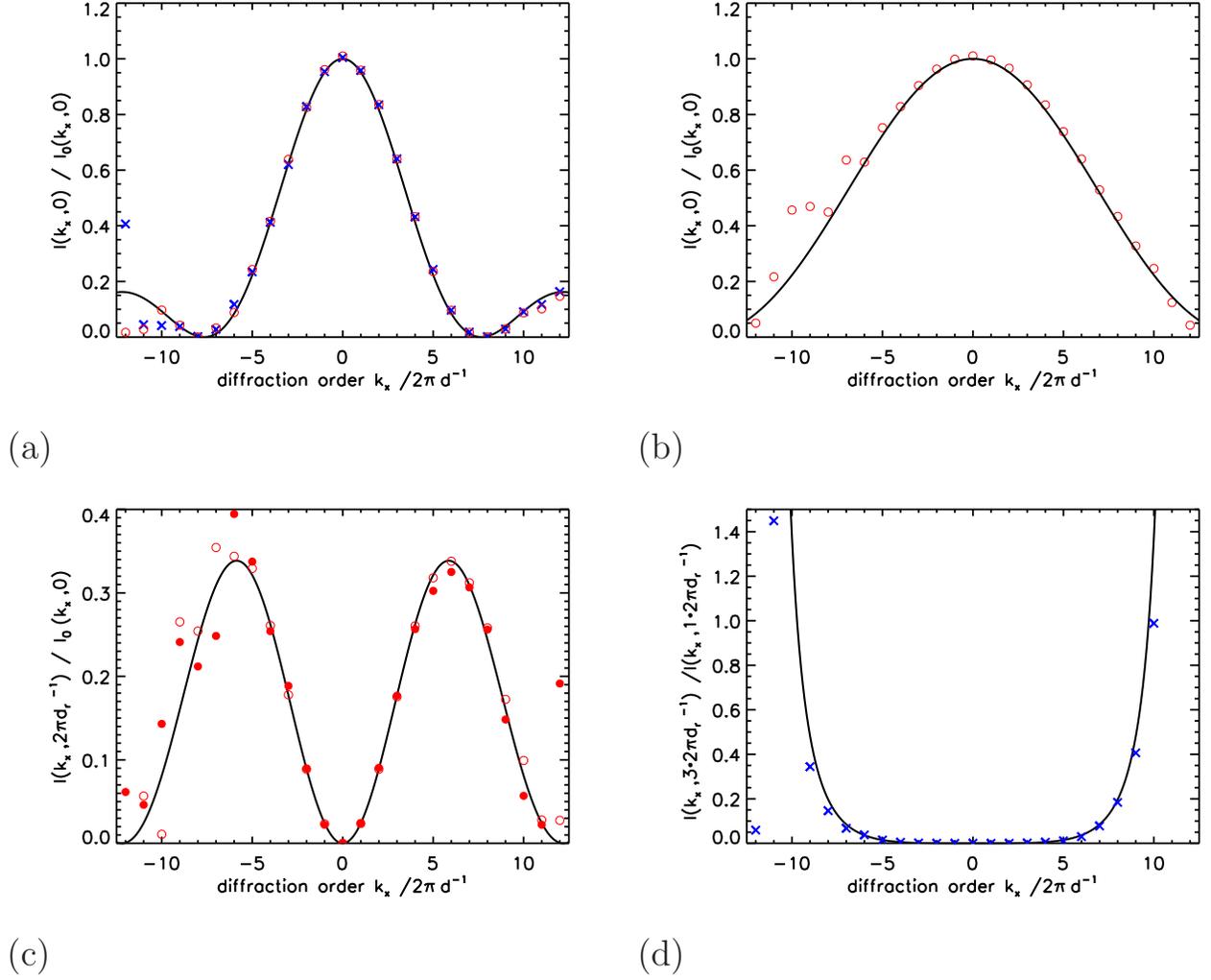}
\end{center}
\caption[schemes]{\label{ratios_fem}  Black curves: Bessel squared functions according to Eqs. (\ref{ler_quotient}), (\ref{lwr_quotient1}), (\ref{lwr_quotient2}).
Data points: ratios of the diffraction intensities obtained from FEM
simulations with respect to the FEM simulation of the undisturbed structure,
denoted by $I_{0}$: LER (red circles) and LWR (blue crosses). 
The damping ratio for the 0$^{\mathrm{th}}$ order in $y$ is shown in (a) for LER and LWR at $a=10$~nm
and in (b) for LER at $a=5$~nm both with $d_{r}=600$~nm. (c) First order in $y$ for
LER at amplitude $a=10$~nm and $d_{r}=300$~nm (closed circles) or \mbox{600 nm} (open circles), respectively. 
(d) Ratio of third and first order in $y$ for LWR at \mbox{$a=10$ nm} and $d_{r}=\mbox{600 nm}$.}
\end{figure}

Figs. \ref{3dfem} (a) and (b) show the diffraction intensities calculated by the FEM simulation. According to the analytical studies of the previous sections, the major difference between the two roughness models should be observed in $I\left(0,\pm 2\pi/d_{r}\right)$. This is shown in more detail in figs. \ref{3dfem} (c) and (d) where numerical values are presented for the $0^{\mathrm{th}}$, $1^{\mathrm{st}}$, and $2^{\mathrm{nd}}$ order in $y$. In the case of LER, a detectable diffraction order for $k_{x}=0$ exists only for $k_{y}=0$, because the Bessel function in Eq. (\ref{ler}) vanishes for all other orders in $y$. In the case of LWR, Eq. (\ref{lwr_kx_0}) states that there are three observable diffraction orders for $k_{x}=0$, which are the 0th and the $\pm$1st order. This is fully consistent with the simulation results shown. In the case of LER, the calculated reflectivity in the orders $(0,\pm1)$ is below  $1\times10^{-6}$, whereas there is a detectable reflectivity of around $3\times10^{-3}$ in the case of LWR, cf. also Figs. \ref{3dfem} (c) and (d). The intensity distributions in the even orders as given by Eqs. (\ref{ler}) and (\ref{lwrgeradeyordnungen}) in $y$ are practically identical for both LER and LWR as can be seen for the 0th and 2nd order, respectively, in Figs. \ref{3dfem} (c) and (d).
The analytical results of Eqs. (\ref{ler_quotient}), (\ref{lwr_quotient1}), (\ref{lwr_quotient2}) are also consistent with the rigorous calculation.
Fig. \ref{ratios_fem} (c) illustrates that, also in the FEM simulation, the
ratio of the first order roughness induced scatter $I\left(k_{x},\pm
2\pi/d_{r}\right)$ and the undisturbed in-plane diffraction $I\left(k_{x},0\right)$ does not depend on the roughness period $d_{r}$. The dependence on the roughness amplitude $a$ is illustrated in Figs. \ref{ratios_fem} (a) and (b).
For even orders in $y$, the effect of line roughness is the same for both LER and LWR, cf. Fig. \ref{ratios_fem} (a). Even the ratio for different odd orders as given by Eq. (\ref{lwr_quotient2}) has been confirmed by the simulation, cf. Fig. \ref{ratios_fem} (d).

\begin{figure}[htbp]
\begin{center}
\includegraphics[width=\textwidth]{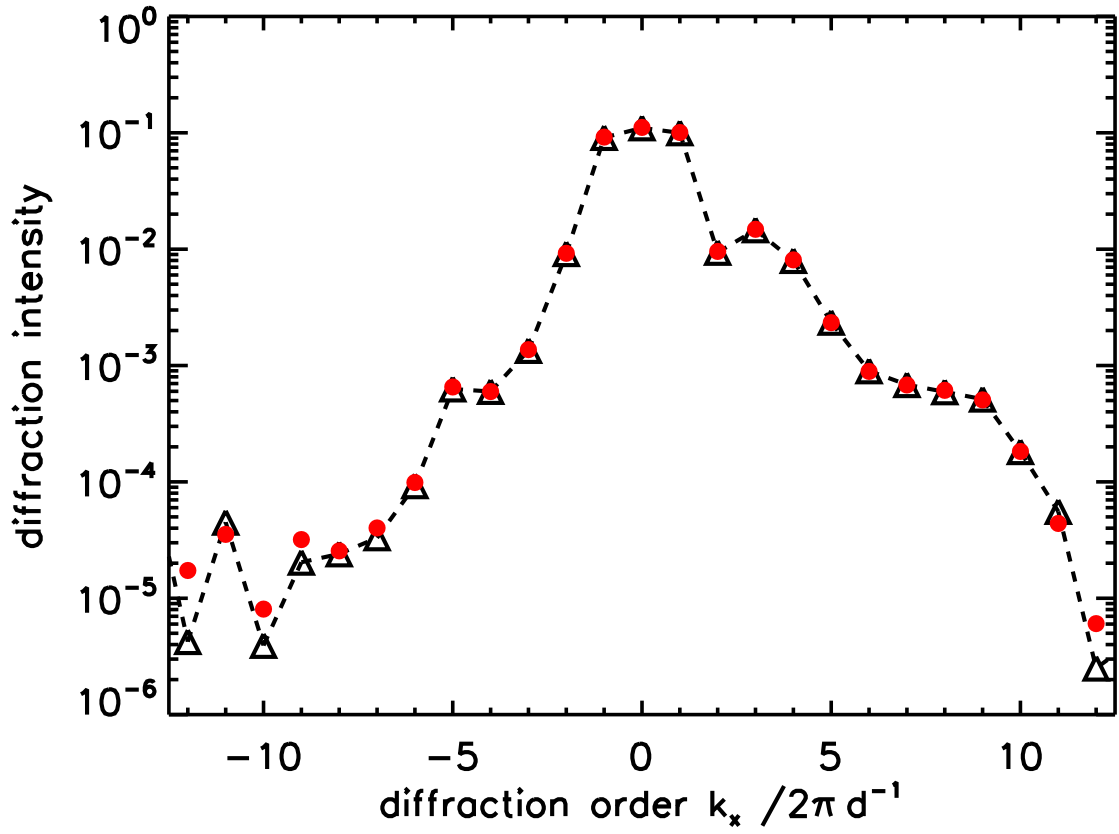}
\end{center}
\caption[]{\label{ler_integrationuebery} The black polygonal chain connects the diffraction intensities of the undisturbed grating (triangles). The red points represent the sum of the LER-disturbed intensities over the orders along the $y$-direction. LER with \mbox{$a=10$ nm}, \mbox{$d_{r}=600$ nm}.}
\end{figure}

Fig. \ref{ler_integrationuebery} illustrates Eq. (\ref{energieerhaltung}). The undisturbed in-plane diffraction intensities correspond to the sum of the disturbed intensities over $y$. Thus, LER scatters the light in the $y$-direction by conserving the energy for each order in $x$.

According to Eq. (\ref{lwr_kx_0}), we obtain

\begin{equation}\label{frac_c_a}
\frac{I_{g}\left(0,\pm2\pi/d_{r}\right)}{I_{g}\left(0,0\right)}=\left(\frac{a}{c}\right)^2
\end{equation}

in the case of LWR. 
As mentioned before, the Fraunhofer approximation for $I_{r}$ does not represent well the real intensities. Particularly, the $0^{\mathrm{th}}$ order $I(0,0)$ is subject to phase effects caused by the real 3D structure of the sample as compared to the 2D approximation of the Fraunhofer diffraction. These effects are, however, present in measured data and accounted for in the FEM calculations. In our case, $I(0,0)$ is suppressed in densely patterned fields of the photomask as compared to the Fraunhofer approximation. A correction factor can be found by normalizing the zeroth order with respect to the first oder in-plane diffraction found by the Fraunhofer model and by the FEM simulation as well.

\begin{eqnarray}
\mathrm{const}&=&\frac{2\tilde{I}_{r}\left(0,0\right)}{\tilde{I}_{r}\left(2\pi/d,0\right)+\tilde{I}_{r}\left(-2\pi/d,0\right)} \times \left[\frac{2I_{r}\left(0,0\right)}{I_{r}\left(2pi/d,0\right)+I_{r}\left(-2pi/d,0\right)}\right]^{-1}\nonumber\\
&=&\frac{2\tilde{I}_{r}\left(0,0\right)}{\tilde{I}_{r}\left(2\pi/d,0\right)+\tilde{I}_{r}\left(-2\pi/d,0\right)}
\times \left( \frac{\sin\left(\pi\mathrm{c}/d\right)}{\pi\mathrm{c}/d}\right)^{2}\nonumber\\
&=&0.468\,.\label{korrekturfaktor}
\end{eqnarray}

\begin{figure}[htbp]
\begin{center}
\includegraphics[width=\textwidth]{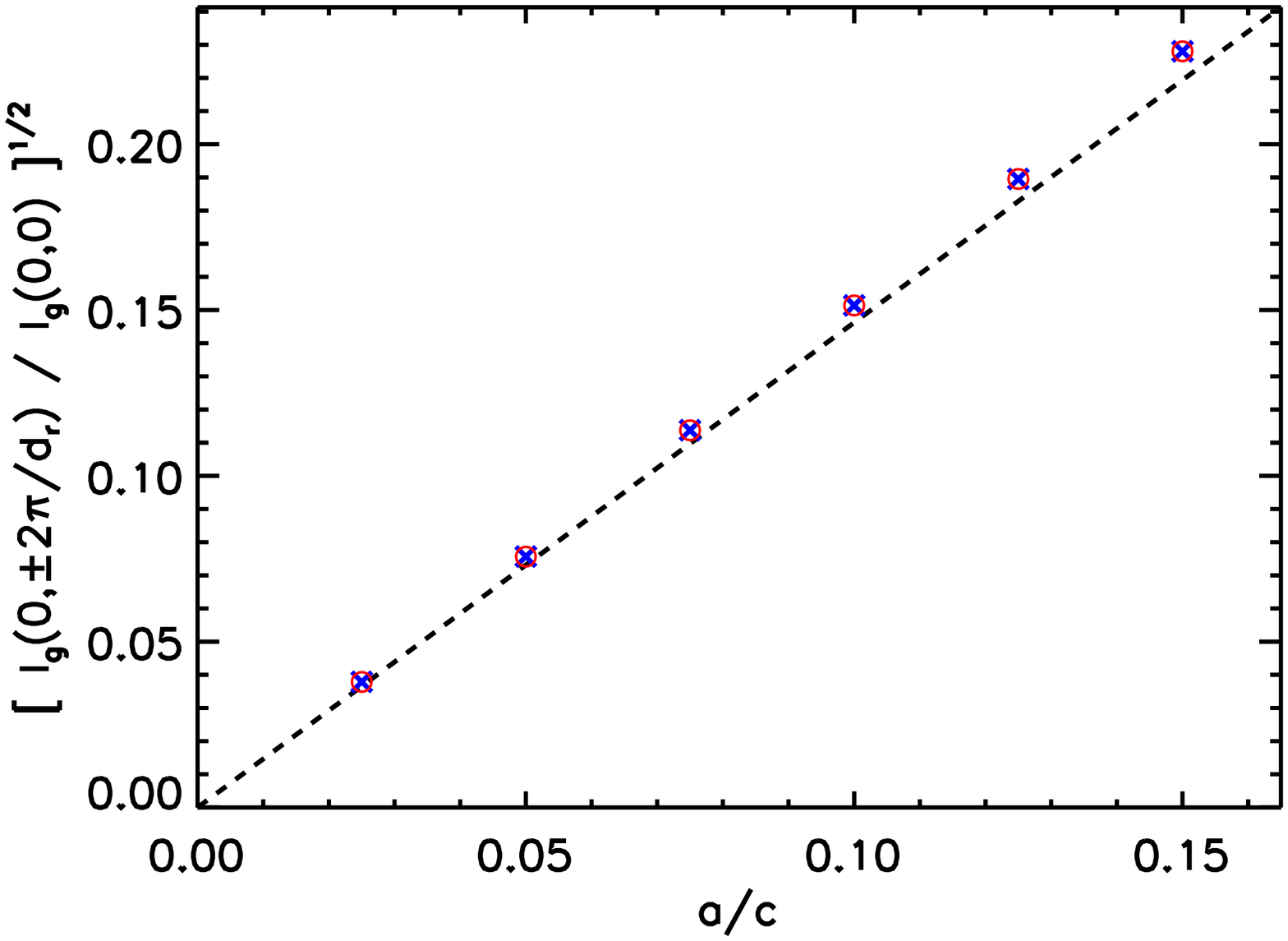}
\end{center}
\caption[]{\label{lwr_xgleichnull_2}  FEM results for $k_{y}=\pm2\pi/d_{r}$ at
$k_{x}=0$ in the case of LWR. Shown are the ratios of the orders +1 (red
circles) and -1 (blue crosses) in $y$ to the $0^{\mathrm{th}}$ order $\tilde{I}_{g}(0,0)$ for 6
amplitudes $a$ at $c=100$~nm. The dashed line shows the slope
$\sqrt{1/\mathrm{const}}$ obtained from the correction factor derived in Eq. (\ref{korrekturfaktor}) to account for the difference between $\tilde{I}_{g}(0,0)$ and $I_{g}(0,0)$ as obtained by FEM and the Fraunhofer approximation, respectively.}
\end{figure}

Here, the second term uses the diffraction intensity calculated by the
analytical Fraunhofer model $I_{r}$ (Eq. (\ref{i_0intensitaet})) and $\tilde{I}_{r}$ is the intensity of the 
undisturbed reference grating calculated by FEM. This correction factor can be used to estimate the ratio $a/c$ from the diffraction orders for $k_{y}=0$ and $k_{y}=\pm2\pi/d_{r}$ at $k_{x}=0$:

\begin{equation}
\left(\frac{a}{c}\right)^2 =\mathrm{const}\times\frac{\tilde{I}_{g}\left(0,\pm2\pi/d_{r}\right)}{\tilde{I}_{g}\left(0,0\right)}\,,
\end{equation}

where $\tilde{I}_{g}$ stands for the simulated and potentially measured intensity of the LWR disturbed grating. 
The result for the simulated intensities as used here is shown in Fig. \ref{lwr_xgleichnull_2}.

\section{Conclusion}

We have studied the influence of line edge and line width roughness on the diffraction intensities of line gratings, e.g. a scatter field of a photomask. For sinusoidal roughness models, the analytical model of Fraunhofer optics leads to a description of the disturbed 2D diffraction pattern by means of Bessel functions of the first kind. Also rigorous simulations of the same structures were performed using a finite-element-based Maxwell solver. Comparing the ratio of the out-of-plane, roughness induced diffraction intensities to the in-plane diffraction of the undisturbed structure as calculated by FEM, it could be shown that the ratio fits very well the Bessel squared functions, as predicted by Fraunhofer optics. For LER and at least for the even off-plane orders caused by LWR, the disturbed diffraction intensities can therefore be estimated using the numerical result of an undisturbed line and space grating, and the results of the Fraunhofer model without explicitely simulating the line roughness by FEM. Some general properties of the 2D diffraction at rough gratings are discussed. In the case of LER, the scattering parallel to the lines occurs by conserving the energy in each order perpendicular to the lines. For the zeroth order, there are no roughness induced out-of-plane scatter intensities for LER and only first order diffraction intensities in the case of LWR.

\section*{Acknowledgments}

The work presented here is part of the CDuR32 project funded by the German Federal Ministry of Education and Research and of the EMRP Joint Research Project IND17.
The EMRP is jointly funded by the EMRP participating countries within EURAMET and the European Union.

\clearpage
 
\clearpage

\clearpage

\clearpage

\clearpage

\clearpage

\begin{thebibliography}{99}

\bibitem{germer07a}
T. A. Germer, ``Effect of line and trench profile variation on specular and diffuse reflectance from a periodic structure,'' \josaa \textbf{24}, 696--701 (2007).

\bibitem{germer07}
T. A. Germer, ``Modeling the effect of line profile variation on optical critical dimension metrology,'' \pspie \textbf{6518}, 65180Z (2007).

\bibitem{naulleau05}
P. P. Naulleau, ``Effect of mask-roughness on printed contact-size variation in extreme-ultraviolet lithography,'' \ao \textbf{44}, 183--189 (2005).

\bibitem{gross09a}
H. Gross, A. Rathsfeld, F. Scholze, and M. B{\"a}r, ``Profile reconstruction in extreme ultraviolet ({EUV}) scatterometry:
  modeling and uncertainty estimates,'' Meas. Sci. Technol. \textbf{20}, 105102 (2009).

\bibitem{boher05}
P. Boher, J. Petit, T. Leroux, J. Foucher, Y. Desi\`{e}res, J. Hazart, and P. Chaton, ``Optical {F}ourier transform scatterometry for {LER} and {LWR}
  metrology,'' \pspie \textbf{5752}, 192--203 (2005).

\bibitem{wang07}
C. Wang, R. L. Jones, E. K. Lin, W. Wu, B. J. Rice, K. Choi, G. Thompson, S. J.
  Weigand, and D. T. Keane, ``Characterization of correlated line edge roughness of nanoscale line
  gratings using small angle x-ray scattering,'' J. Appl. Phys. \textbf{102}, 024901 (2007).

\bibitem{scholze08}
F. Scholze and C. Laubis, ``Use of {EUV} scatterometry for the characterization of line profiles and line roughness on photomasks,'' \pspie \textbf{6792}, 67920U (2008).

\bibitem{bodermann09}
B. Bodermann, M. Wurm, A. Diener, F. Scholze, and H. Gro{\ss}, ``{EUV} and {DUV} scatterometry for {CD} and edge profile metrology on
  {EUV} masks,'' \pspie \textbf{7470}, 74700F (2009).

\bibitem{pomplun08}
J. Pomplun, S. Burger, F. Schmidt, F. Scholze, C. Laubis, and U. Dersch, ``Metrology of {EUV} masks by {EUV}-scatteometry and finite element
  analysis,'' \pspie \textbf{7028}, 70280P (2008).

\bibitem{burger11}
S. Burger, L. Zschiedrich, J. Pomplun, and F. Schmidt, ``Rigorous simulations of 3{D} patterns on extreme ultraviolet
  lithography masks,'' \pspie \textbf{8083}, 80831B (2011).

\bibitem{scholze11}
F. Scholze, B. Bodermann, H. Gro{\ss}, A. Kato, and M. Wurm, ``First steps towards traceability in scatterometry,'' \pspie \textbf{7985}, 79850G (2011).

\bibitem{kato11}
A. Kato and F. Scholze, ``The effect of line roughness on the diffraction intensities in angular resolved scatterometry,'' \pspie \textbf{8083}, 80830K (2011).

\bibitem{kato10}
A. Kato and F. Scholze, ``Effect of line roughness on the diffraction intensities in angular resolved scatterometry,'' \ao \textbf{49}, 6102--6110 (2010).

\bibitem{gross11}
H. Gross, M.-A. Henn, A. Rathsfeld, and M. B{\"a}r, ``Stoachstic modeling aspects for an improved solution of the inverse problem in scatterometry,'' in \emph{Advanced Mathematical and Computional Tools in Metrology and Testing IX, Series on Advances in Mathematics for Applied Sciences} \textbf{84} 202-209 (2012).

\bibitem{schuster08}
T. Schuster, S. Rafler, K. Frenner, and W. Osten, ``Influence of line edge roughness ({LER}) on angular resolved and on
  spectroscopic scatterometry,'' \pspie \textbf{7155}, 71550W (2008).

\bibitem{Bergner10}
B. C. Bergner, T. A. Germer, and T. J. Suleski, ``Effective medium approximation for modeling optical reflectance from gratings with rough edges,'' J. OPt. Soc. Am. A \textbf{27}, 1083-1089 (2010).

\bibitem{schuster09}
T. Schuster, S. Rafler, V. Ferreras Paz, K. Frenner, and W. Osten, ``Fieldstitching with {K}irchhoff-boundaries as a model based
  description for line edge roughness ({LER}) in scatterometry,'' Microelectronic Eng. \textbf{86}, 1029--1032 (2009).

\bibitem{colton98}
D. Colton, R. Kress, ``Inverse acoustic and electromagnetic scattering theory,''
             Applied Mathematical Sciences \textbf{93}, Springer Berlin (1998).

\bibitem{pomplun07pssb}
J. Pomplun, S. Burger, L. Zschiedrich, and F. Schmidt, ``Adaptive finite element method for simulation of optical nano structures,'' Phys. Stat. Sol. B \textbf{244}, 3419-3434 (2007).

\bibitem{lockau11pw}
D. Lockau, L. Zschiedrich, S. Burger, F. Schmidt, F. Ruske, and B. Rech, ``Rigorous optical simulation of light management in crystalline
  silicon thin film solar cells with rough interface textures,'' \pspie \textbf{7933}, 79330M (2011).

\bibitem{schaedle07dd}
A. Sch{\"a}dle, L. Zschiedrich, S. Burger, R. Klose, and F. Schmidt, ``Domain decomposition method for {M}axwell's equations: {S}cattering
  off periodic structures,'' J. Comput. Phys. \textbf{226}, 477--493 (2007).

\bibitem{burger11pm}
S. Burger, L. Zschiedrich, J. Pomplun, F. Schmidt, A. Kato, C. Laubis, and F. Scholze, ``Investigation of {3D} patterns on {EUV} masks by means of
  scatterometry and comparison to numerical simulations,'' \pspie \textbf{8166}, 81661Q (2011).

\end{thebibliography}
\end{document}